# Leveraging Google Earth Engine platform to characterize and map small seasonal wetlands in the semi-arid environments of South Africa

Siyamthanda Gxokwe*, Timothy Dube, Dominic Mazvimavi

Institute for Water Studies, Department of Earth Science, University of the Western Cape, Private Bag X17, Bellville, 7535, Cape Town, South Africa

*Corresponding author's email: 3050512@myuwc.ac.za

## Abstract

Small seasonal wetlands are a distinctive feature of the semi-arid environments. Although these systems are small (<10 ha – 2500 ha) and ephemeral in nature, they provide habitats to aquatic flora and fauna, and play a critical role in sustaining livelihoods through the provision of ecological services. There is a great concern that these wetlands are not routinely monitored, which has contributed to their poor management. So far, scientific research strides have been noted in mapping and understanding the spatial coverage and ecohydrological dynamics of wetlands in semi-arid areas using broadband multispectral data. However, because of the sensing characteristics of these platforms, mapping and monitoring of small wetlands remains a challenge. This is further hampered by the lack of robust and effective satellite data processing techniques. The recent advancement in data analytics tools such as the introduction of the Google Earth Engine (GEE) platform provides opportunities to improve the mapping and understanding of these small and scattered wetland systems, a previously daunting task with conventional methods. This study sought to assess the capabilities of GEE cloud-computing platform in characterising and mapping small and seasonal flooded wetlands at site specific scale using the new generation Sentinel 2 composite data. Specifically, a multi- year (2016-2020) Sentinel-2 composite data were used to map two wetlands with varying spatial extents (i.e., Lindani valley bottom and Nylsvley floodplain covering an area of 28 ha and 1369 ha, respectively). Wetland classification was executed using the Object based Random Forest (RF), Support Vector Machine (SVM), Classification and Regression Tree (CART) and Naïve Bayes (NB) in GEE. The results demonstrated the capabilities of using the Google Earth Engine platform to characterize and map seasonally flooded wetlands in semi-arid environments with acceptable accuracy. The CART, RF, SVM were found to be superior to NB, which had the lowest overall accuracy. The findings of this study underscore the relevance of GEE and Sentinel 2 composite data in characterizing and mapping small and seasonal wetlands found in semi-arid environments.

## 1. Introduction

Wetlands play a critical role in the hydrological cycle, sustaining livelihoods and aquatic life and biodiversity. They occupy transition zones between the aquatic and terrestrial environments and share the characteristics of both zones (Gxokwe *et al*., 2020). Wetlands cover a proportion of about 4-6% of the global land surface and are ranked amongst the very diverse ecosystems on earth (Mahdianpari *et al.*, 2019). Semi-arid areas are dominated by small seasonally or intermittently flooded wetlands, and the flooded area depends on the balance between precipitation and evapotranspiration (Ruiz, 2008). As such, semi-arid wetlands tend to host more invertebrates than the permanently inundated systems because of their oxygenated period resulting from the episodic inundation. Although wetlands in the semi-arid areas host most invertebrates that would not survive in surrounding landscape, their conservation status is still not prioritised (Chen and Liu, 2015). This is mainly due to their size and ephemeral nature, which therefore result in these wetlands being neglected in monitoring and management programs, thus leading to the loss of inherent ecosystems goods and services provision (Li *et al.*, 2015). Globally, the abundance and quality of wetlands in semi-arid environments is reported to be declining due to climate change and variability as well as poor land management practices (Mahdianpari *et al.*, 2019). Gebresllassie *et al.*, (2014) reported a significant wetland loss in the semi-arid Ethiopia due to lack of policies to safeguard these systems thus resulting in loss of socio-economic services. In semi-arid parts of China, it has been reported that about 30% of wetlands have been lost over the past 50 years due to anthropogenic activities, with most disappearing between 1990 and 2000 (Liu *et al.*, 2017). In South Africa, it is reported that over 50% of wetlands have been eradicated in some catchments due to climate change and anthropogenic activities (Day *et al.*, 2010). Given the significance of ecological services provided by wetlands, it is imperative that these systems are sustainably managed. The basis for sustainable management of wetlands is frequent monitoring of their ecohydrological dynamics to produce consistent and comparable information, which is lacking in most semi-arid regions, especially in sub-Saharan Africa (Mahdianpari *et al.*, 2019). The availability of Earth Observation (EO) data offers a platform to map and monitor wetlands in a spatially explicit manner in different climatic zones and regions, and where the monitoring systems are not available (Gxokwe *et al.*, 2020). The challenge however is that these data come in a range of spatial, spectral and temporal resolutions which in most cases present difficulties when mapping semi-arid wetlands to the finest detail, especially with the use of coarse to medium resolution dataset such as Moderate Resolution Imaging Spectroradiometer (MODIS). Wetlands in the semi-arid regions are mostly heterogeneous, with no definitive boundaries, and are spectrally similar to the surrounding landscapes. This therefore result in difficulties when separating these systems from the surrounding landscapes using coarse to medium spatial resolution data (Mahdianpari *et al.*, 2020). High resolution data such as Wordview-2 and Satellite Pour I'Obeservation de la Terre (SPOT 6-7) are mostly commercial and require complex processing algorithms and are therefore not feasible to monitor the spatial characteristics of semi-arid wetland over large areas and overtime (Gxokwe *et al*., 2020). Advancements in data analytic tools and platforms provide new opportunities for strengthening research on wetlands monitoring and assessment across various scales, with development of cloud computing platforms such as Microsoft Azure (MA), Amazon web services (AWS) and Google earth engine (GEE), amongst others. The AWS was launched in 2006 and contains several remote sensing data ranging from Sentinel-1, Sentinel-2, Landsat 8, and National Oceanographic and Atmospheric Administration (NOAA), Advanced High-Resolution Rapid Refresh (HRRR) Model (Tamiminia *et al*., 2020). Although AWS provides benefits of accessing a large suite of machine learning algorithms and artificial intelligence, the platform offers pay-as-you go services. Microsoft Azure was launched in 2010 for building, deploying and managing applications and services through Microsoft-managed data centres. The platform consists of advanced machine learning algorithms, Landsat and Sentinel-2 data from 2013 to present for only North America as well as MODIS data from 2000 to present. The recently introduced GEE platform offers a parallelised processing on Google cloud that makes it easy to process a stack of images at once rather than relying on a single date image. The platform also consists of 40 years petabyte scale of pre-processed remotely sensed data, which include Landsat, MODIS, (NOAA AVHRR), Sentinel 1, 2, 3 and 5-P; and Advanced Land Observing Satellite (ALOS) data as well as advanced machine learning algorithms. The other data type that are available on GEE platform include climate and geophysical data as well as ready to use products such as Normalised Difference Vegetation Index (NDVI) and Enhanced Vegetation index (EVI). Although GEE was launched over a decade ago, its application to remote sensing of wetlands including the small seasonal flooded systems in the semi-arid regions is still limited. A review by Tamiminia *et al*., (2020) reported 13 wetlands and mangroves studies that utilized GEE platform between 2010 and 2019 in all climatic zones, with most of these studies exploiting Landsat 8 data. The review by Kumar and Mutanga, (2018) also reported that of 8% (out of the 300 articles identified) of the studies utilized the GEE platforms for wetlands and hydrological related research globally. The most recent studies by Mahdianpari *et al.,* (2019, 2020) used GEE to monitor semi-arid wetlands in Canada with reasonable accuracies (70% - 90%), however, the studies focused on large-scale mapping. There is therefore a need to explore the capabilities of this platform for mapping and determining characteristics of semi-arid wetlands. Owing to that background, the overarching goal of this study was to characterize and map two seasonal flooded wetlands in the semi-arid

Limpopo Transboundary River Basin in South Africa using GEE cloud-computing platform. Specifically, the objectives were to; (1) evaluate the capabilities of GEE cloud-computing platform in producing customized wetland cover maps at reasonable accuracy, using the high-resolution Sentinel-2 data, and (2) identify a suitable GEE machine learning algorithm for accurately detecting and mapping semi-arid seasonal wetlands characteristics using multi-year Sentinel 2 composite data.

## 2. Study area

Two wetlands located in the Limpopo Transboundary River Basin in South Africa (LTRB) were monitored (Figure 1). The LTRB is shared by four countries namely South Africa, Botswana, Mozambique and Zimbabwe, and covers an area of 412 000 km$^2$ (Mosase *et al*., 2019). The basin has a semi-arid climate, with wet summers and dry winters. The Mean Annual Precipitation (MAP) in the LTRB ranges from 300 to 700 mm/year with most of the rainfall occurring during the October to April period (Botai *et al.*, 2020). Temperatures in the LTRB follows a distinct seasonal cycle with coolest months in winter (June-August), and hottest in the late summer months (late November- early December). The mean daily temperatures in LTRB can go up to 40°C (Sawunyama et al., 2006). The mean annual evaporation varies between 1600 mm/yr and 1700 mm/yr in the southeastern mountainous region, and from 2600 mm/yr to 3100 mm/yr in the western and central regions (Sawunyama *et al*., 2006). The two studied seasonal flooded wetlands are the Nylsvley floodplain and Lindani valley bottom in South Africa. Nylsvley floodplain is Ramsar protected system located at 24°39′17″S and 28°41′28″E near Mookgopong and Modemolle towns in the Limpopo Province of South Africa. The wetland forms a 70 km long floodplain along Mogalakwena River, which is a tributary of the Limpopo River (Dzurume, 2021). The dominant vegetation species occurring in the Nylsvley floodplain include the common grass species such as *Oryza longistaminata* (rice grass) and *Phragmites australis* (common reeds), and tree species such as *Acacia tortilis, Acacia nilotica* and *Acacia karoo*. The Nylsvley floodplain receives most of the inflows from the seasonal rivers such as Olifantsspruit, Groot and Klein Nyl (Dzurume, 2021). Although the floodplain is 70 km long, the study focuses on the 13.69 km$^2$ portion within the boundaries of the Nyls Nature Reserve , which was accessible during the study period. Lindani valley bottom wetland is located between 24°03′01.36″S and 28°41′43.37″S within the boundaries of the Lindani Private Game lodge at Vaalwater Limpopo province of South Africa. The wetland covers an area of about 28 ha and mostly receives water from the rainfall and groundwater seeps from several springs in the area. The dominant vegetation species include the *Oryza* *longistaminata* (rice grass), *Phragmites australis* (common reeds), *Scirpoides dioecus* (Kunth) as well as *Cynodon dyctolon* (Bermuda grass).

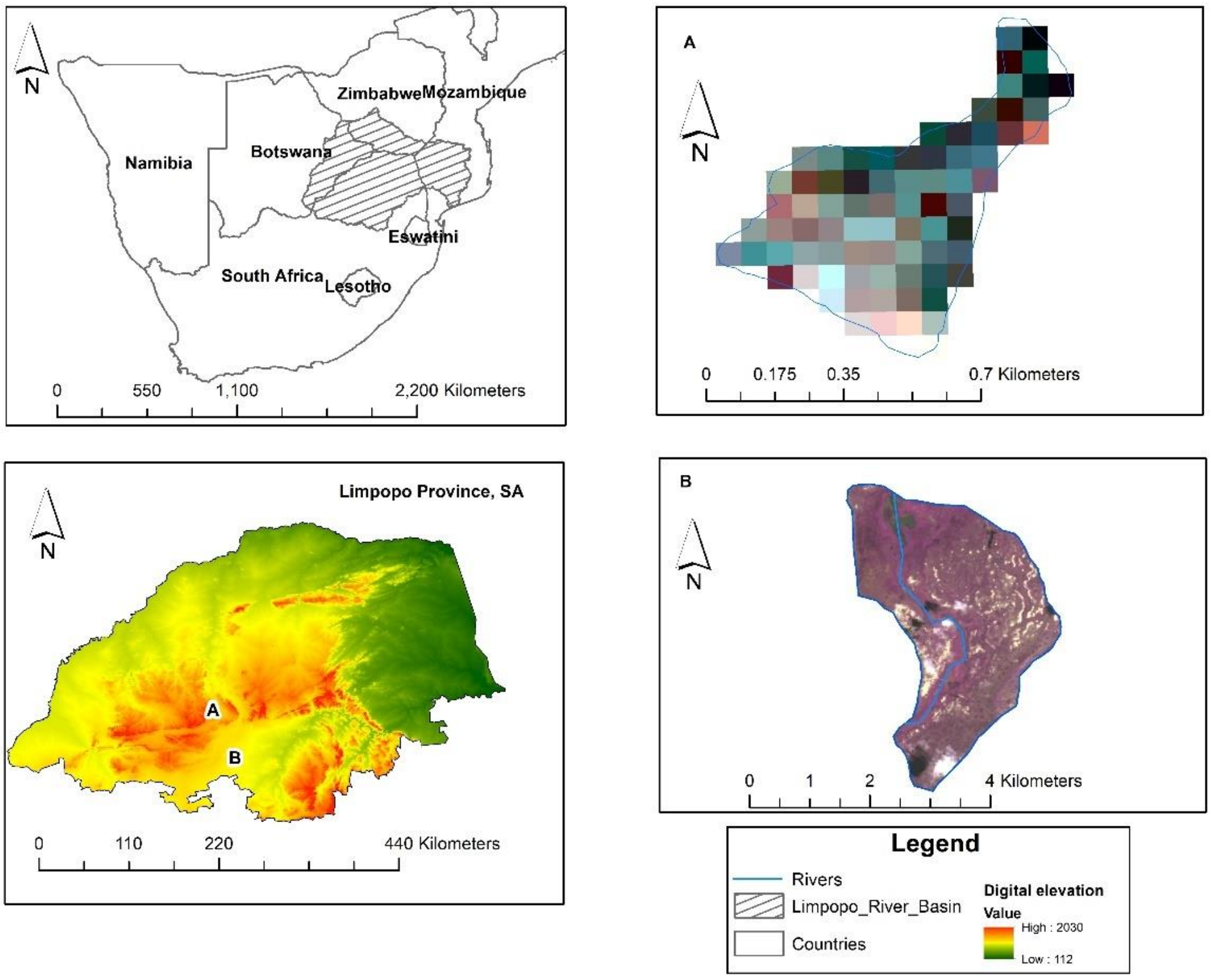

**Figure.1.** Location of the studied wetlands, a) Lindani valley bottom b) Nylsvley floodplain within the Limpopo Transboundary River Basin.

## 3. Materials and methods

### *3.1 Field data*

Land cover data were collected in the field during the end of dry season and beginning of wet season between 28 September 2020 and 1 October 2020. The data collected included, six hundred ground truth points collected on both wetlands with locations determined used a handheld Geographical Positioning System (GPS). The points were collected, using a stratified random sampling approach, where the wetlands were subdivided into 10 m x 10 m quadrants based on the Sentinel-2-pixel sizes. In each pixel a maximum of 20 sampling points were collected depending on the dominating landcover classes. The collected ground truth points were used for the GEE wetland model training and validating satellite derived wetland cover classes.

### *3.2 Remote sensing data acquisition and processing*

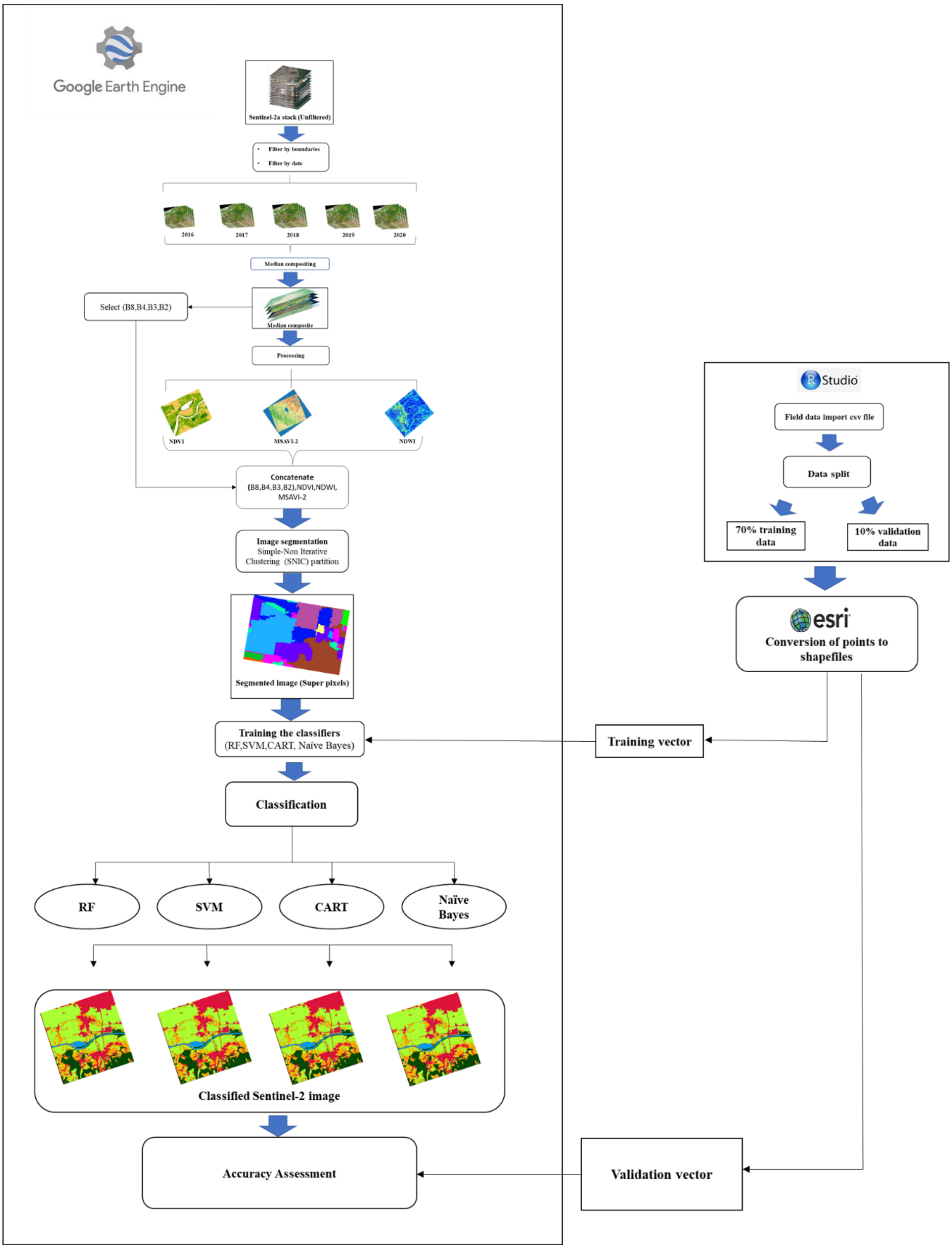

**Figure. 2.** Steps undertaken to characterise and map the two wetlands using GEE.

The acquisition and processing of remotely sensed data in this study were executed following the steps shown on Figure 2. Sentinel-2A pre-processed Top of the Atmosphere (TOA) reflectance images were acquired from Google Earth Engine (GEE) database. The image stack was filtered to represent the period of 2016-01-01 to 2020-12-31, and the area within the selected wetland boundaries using the codes ee.Filter.Date () and Image.filterBounds (). The period was chosen because in 2016, drought was reported in the Limpopo Basin, which caused significant changes in most surface water systems including wetlands in the area. The study aimed to establish if drought induced changes within the wetlands could be determined. Two hundred ninety-six images were obtained after the filtering process. The acquired image stack was then normalized for illumination effects (i.e., shades) and minimization of clouds, using median compositing. The median composite works by reducing a stack of images through the calculation of median of all values at each pixel across the stack of all matching bands thus minimising the effects of shades and clouds (Mahdianpari *et al.*, 2019). The median composite in this study was executed using the code Median () on GEE. This was then used to calculate the Normalised Difference Vegetation Index (NDVI), Normalised Difference Water Index (NDWI) and Modified Soil Adjusted Vegetation Index 2 (MSAVI-2), using the equations given in Table 1. The NDVI is one of the most widely used vegetation index in wetlands studies due to the sensitivity to photosynthetically active biomass and can discriminate between vegetation and non-vegetation as well as wetland from non-wetland features (Liu and Huete, 1995). The NDWI was selected due to the sensitivity to open water making enabling the discrimination of water from land surfaces (McFeeters, 1996). MSAVI-2 was chosen to improve the limitations of the NDVI. In addition to the extracted indices, the near infrared, red, green and blue bands were selected and concatenated to the NDVI, NDWI and MSAVI-2 outputs to produce an image with only the bands for wetlands classification. The produced composite was subjected to Object Based Image Analysis (OBIA). OBIA was selected because of the superiority against pixel-based classification as shown in various wetlands mapping studies such as Berhane *et al.*, (2018), Kamal and Phinn, (2011) and Kaplan and Avdan, (2017). Moreover, the approach not only relies on spectral characteristics of each pixel but also considers other pixel characteristics, such as size, shape and contextual information to improve spectral separability within classes of the heterogeneous wetlands (Halabisky, 2011). The first step in OBIA is image segmentation. The process involves partitioning of the image into multiple discreet and non-overlapping segments based on a specific criterion (Dlamini *et al.*, 2021). In this process, individual pixels are merged to produce larger objects. This increases the discrimination of spectrally similar objects using the texture, shape and contextual features and prevents the "salt and pepper" noise in the final classification map (Mahdianpari *et al.*, 2019; Dlamini *et al.*, 2021). In this study, a Simple Non-Iterative Clustering (SNIC) algorithm was used to segment the composite. The SNIC algorithm was chosen because of its simplicity, memory efficiency, processing speed as well as the ability to incorporate connectivity between pixels after the algorithm has been initiated (Achanta and Süsstrunk, 2017). The SNIC algorithm starts the process of image segmentation by initialising the centroids pixels on a regular grid image, then the dependency of each pixel relative to the centroids is established using the distance in five-dimensional space of colour and spatial coordinates. In particular, the distance integrates normalized spatial and colour distances to produce uniform super pixels (Achanta and Süsstrunk, 2017). The candidate pixel is selected based on the shortest distance from the centroid (Achanta and Süsstrunk, 2017). The SNIC algorithm was executed using the code ee.Algorithms.Image.Segmantation.SNIC () on GEE and the output was an image with super pixels, calculated textures, areas, sizes and perimeters for all the super pixels.

**Table 1.** Features extracted from the optical data.

| **Data** | **Data extracted** | **Formula** | **Band Width (nm)** | **Reference** |
|---|---|---|---|---|
| Sentinel-2 | B8 - Near-infrared (NIR) | | 842 | |
| | B4 - Red | | 665 | |
| | B3 – Green | | 560 | |
| | B2 – Blue | | 490 | |
| | Normalized Difference Vegetation Index (NDVI) | $\frac{NIR - Red}{NIR + Red}$ | - | (Liu and Huete, 1995) |
| | Normalized Difference Water Index (NDWI) | $\frac{NIR - Green}{NIR + Green}$ | - | (McFeeters, 1996) |
| | Modified Soil Adjusted Vegetation index (MSAVI-2) | $\frac{2 \times NIR + 1 - \sqrt{(2 \times NIR + 1)^2 - 8 \times (NIR - Red)}}{2}$ | - | (Qi *et al.*, 1994) |

*3.3 Adopted wetland classification scheme*

Satellite image classification was executed, using the Random Forest (RF), Support Vector Machine (SVM), Classification and Regression Tree (CART) and Naïve Bayes (NB) algorithms. The RF is an ensemble classifier, which consists of a combination of tree classifiers, and each classifier is generated using the random factor sampled independently from the input vector data. Each tree cast a unit vote for a popular class to classify an input vector (Pal, 2005; Ao *et al.*, 2019). An advantage of RF algorithm is the ability to handle large differentiations within landcover classes, and noise data can be neutralised (Slagter *et al.*, 2020). The SVM applies a sophisticated kernel function to classify data sets with complex decision surface. One of the strengths of SVM is that the uncertainty in model structure is decreased (Oommen *et al.*, 2008). CART is a tree-based classification algorithm that measures the dependence relation of one variable to other variables (Simioni *et al.*, 2020). NB forms part of the simple probabilistic classifiers based on applying Bayes theorem with strong independence assumption between features (Simioni *et al.*, 2020). The four Algorithms were chosen because of their acceptable performance demonstrated in several wetland studies (Simioni *et al.*, 2020; Slagter *et al.*, 2020; Dlamini *et al.*, 2021). Prior to classification, the algorithms were trained using 70% of the field collected data. The data were randomly split in R-studio to produce 70% training set, and 30% validation set. The code used to split data in R-studio was " wasdt = sort(sample(nrow(data), nrow(data)*.7))". After splitting, the data were converted to GIS files, and then imported to GEE to modelling training and validating. The training of the classifiers was executed, using the code "ee.Classifier.train ()". The classification using the latter algorithms was then implemented on the segmented image using the code "Image.classify()" in GEE.

### *3.4 Accuracy assessment*

In evaluating the performance of the classification algorithms, three evaluation indices were used. These are overall accuracy (OA), producer's accuracy and user's accuracy. In addition, scatter plots as well as Jeffries-Matusita (JM) distances were used to establish the separability of different wetland classes using the selected spectral bands. JM is a parametric criterion that ranges between 0 and 2. This criterion uses distances (Equation.1 & 2) between the class means and the distribution of values from the means to assess the separability of one class from the other (Dabboor *et al.*, 2014; Wang, Qi and Liu, 2018). Distances approaching 2 indicate a greater average distance between two classes and therefore separable using the data type. The OA was used to measure the efficiency of the used algorithms and was quantified as the ratio of total correctly labelled samples and the total number of testing samples. The producer's accuracy was used to measure the probability that the reference sample is correctly classified on the map. The user's accuracy was used as an indicator of the probability that a classified pixel in the wetland cover classification map accurately represent that category on the ground. The accuracy assessments with the latter indices were computed executed in GEE. The JM distance presented in Dabboor *et al.*, (2014) is given as:

$$JM = 2\,(1 - e^{-B}) \tag{1}$$

Where B is the Bhattacharyya distance and quantified as

$$B = \frac{1}{8}(\mu_i - \mu_j)^T \left(\frac{\Sigma_i + \Sigma_j}{2}\right)^{-1} (\mu_i - \mu_j) + \frac{1}{2}\ln\left(\frac{|(\Sigma_i + \Sigma_j)/2|}{\sqrt{|\Sigma_i||\Sigma_i|}}\right) \tag{2}$$

Where $\mu_i$ and $\Sigma_i$ are the mean and covariance matrix of class *i* and $\mu_j$ and $\Sigma_j$ are the mean and covariance of class *j*.

## 4. Results

### *4.1 Class separability*

Scatter plots (Figure. 3) showing spectral reflectance values per class of the studied wetlands indicate that most wetland classes are not distinguishable using B2, B3, and B4 for both wetlands except long grass in the Nylsvley flood plain, which was discernible from the rest of the classes using B4 and B8. The results also indicate that Water class in the Lindani valley bottom has high spectral reflectance in the NIR region although the anticipation was low reflectance in this spectral region.

a)

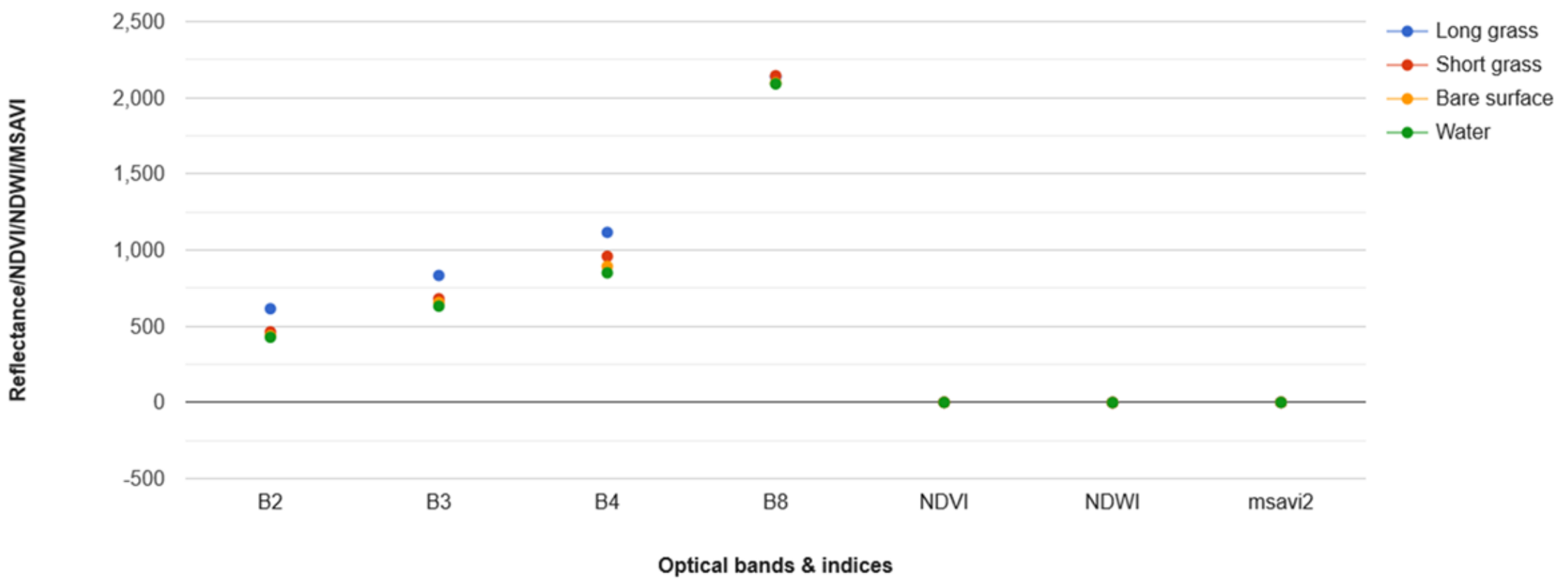

b)

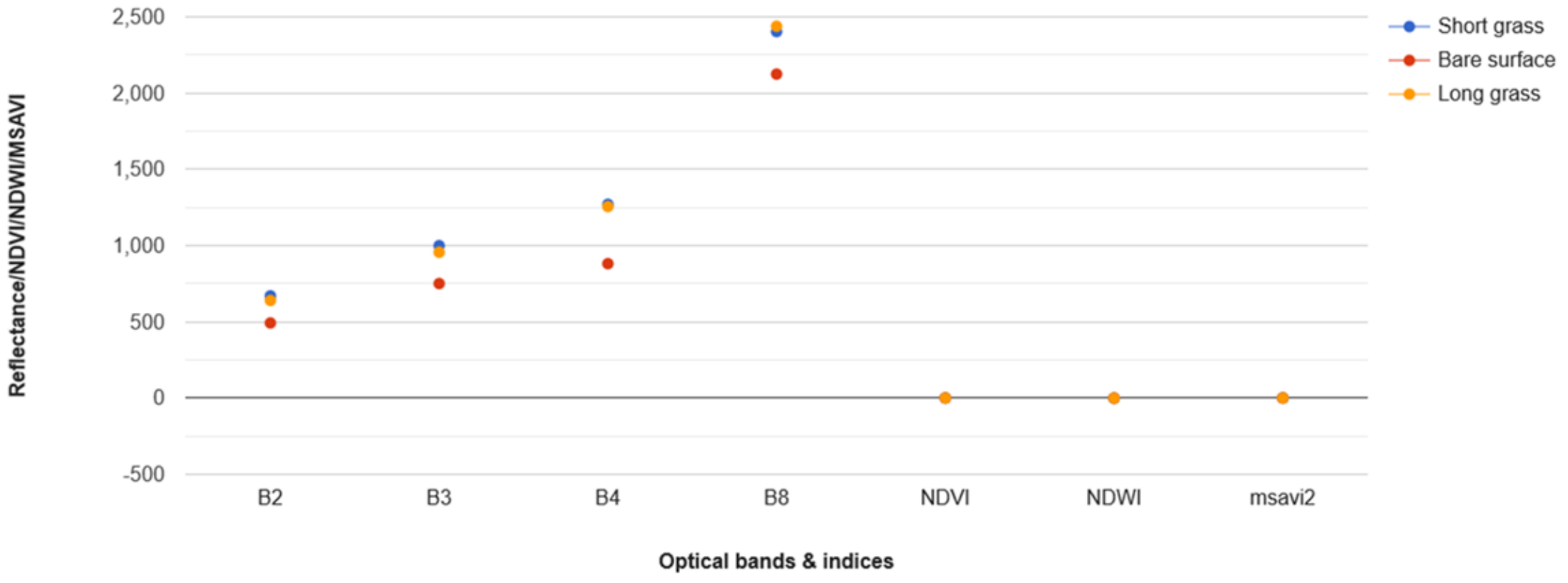

**Figure. 3** Wetland cover class spectral reflectance values for a) Lindani valley bottom and b) Nylsvley floodplain, using pixel values extracted from the training data.

The JM distances obtained from the multi-year median composite (Table 2 and 3), indicate that wetland features were hardly distinguishable from single optical bands for both wetlands. The least distinguishable classes were Bare surface and Water from the Lindani valley bottom as well as Bare surface and Short grass from the Nylsvley floodplain. The JM distances for all the least distinguishable classes were not exceeding 1.4 for both wetlands. The results also show that synergic use of all optical features significantly increased the separability between classes with JM distances exceeding 1.8 for both wetland classes.

**Table 2**. JM distances between wetland cover classes in the Lindani valley bottom

| **Optical features** | **D1** | **D2** | **D3** | **D4** | **D5** | **D6** |
|---|---|---|---|---|---|---|
| **NIR** | 0.0016 | 0.0131 | 0.1036 | 0.0135 | 1.272 | 0.1398 |
| **Red** | 0.3146 | 0.4311 | 0.5635 | 0.0502 | 0.1576 | 0.0314 |
| **Green** | 0.4331 | 0.6508 | 0.8978 | 0.0600 | 0.2714 | 0.0966 |
| **Blue** | 0.4675 | 0.6381 | 0.8861 | 0.0418 | 0.2193 | 0.0858 |
| **NDVI** | 0.2728 | 0.3044 | 0.6300 | 0.0385 | 0.1823 | 0.1127 |
| **NDWI** | 0.3478 | 0.4028 | 0.4346 | 0.0036 | 0.0168 | 0.0089 |
| **MSAVI2** | 0.3124 | 0.3509 | 0.6849 | 0.0041 | 0.1948 | 0.1187 |
| **ALL** | 2 | 2 | 2 | 2 | 2 | 2 |

*** D1: Long grass & Short grass, D2: Long grass & Bare surface, D3: Long grass & Water, D4: Short grass & Water, D5: Short grass and Water, D6: Bare surface and Water.*

**Table 3**. JM distances between wetland cover classes in the Nylsvley floodplain

| **Optical features** | **D1** | **D2** | **D3** |
|---|---|---|---|
| **NIR** | 0.2586 | 0.3839 | 0.0342 |
| **Red** | 0.6208 | 0.5998 | 0.0016 |
| **Green** | 0.6913 | 0.5289 | 0.0226 |
| **Blue** | 0.7211 | 0.5828 | 0.0251 |
| **NDVI** | 0.5663 | 0.5534 | 0.0358 |
| **NDWI** | 0.6675 | 0.3876 | 0.1351 |
| **MSAVI2** | 0.5986 | 0.0038 | 0.0524 |
| **ALL** | 2 | 2 | 1.990 |

***D1: Short grass & Long grass, D2: Long grass & Bare, D3: Bare surface & Short grass.*

*4.2 Classification results and accuracies*

Four GEE algorithms were applied to a median composite of Sentinel-2 images to produce custom maps for two seasonal flooded wetlands of variable sizes in the LTRB. Fig. 7 and 8 show the custom maps for the studied wetlands. Overall Accuracies based on the algorithms used ranged between 20% and 80% for both wetlands, with Random Forest (RF) having high OA for both Lindani valley bottom and Nylsvley floodplain (68.8% and 80.55%) and Naïve Bayes (NB) having the low OA values for both wetlands (29.5% and 25%) (Figure 4). The other algorithms had reasonable accuracy with the Support Vector Machine (SVM) attaining 66.6% for Lindani and 62.29% for Nylsvley. The Classification and Regression Tree (CART) achieved an OA of 62.30% for Lindani and 75% for Nylsvley, thus proving the superiority of RF among other algorithms used in this study.

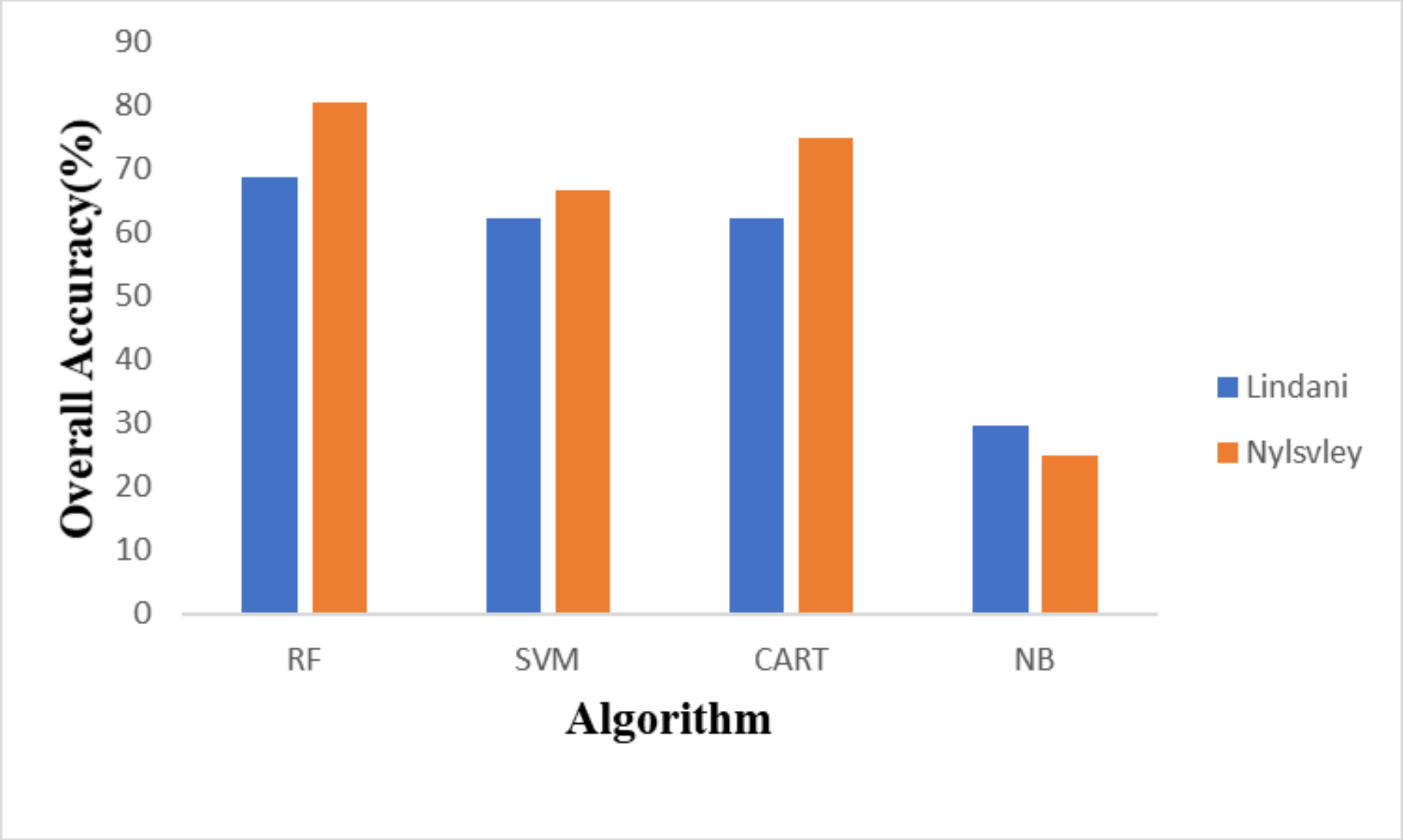

**Figure .4** Overall Accuracy comparison between the algorithms used.

Figures 7 and 8 show the distribution of wetland cover classes at 10 m spatial resolution. The maps illustrate fine separation between the wetland classes for all the algorithms. However, for the Nylsvley floodplain, the water class could not be detected due to the unavailability of training data representing the class. The areas of the various land cover types in Figure 5 show that based on all the algorithms except NB the most dominating class in the Lindani valley bottom is Short grass consisting of *Cynodon dyctolon* and *Oryza longistaminata*. In contrast, the NB identified water as being the most dominant class. The area of the short grass ranged between 5 ha and 25 ha with the NB model identifying the smallest area for this cover type. The producer's accuracy and user's accuracy for the dominating short grass ranged between 20% and 91%, with NB model having the lowest producer's accuracy (Figure 6), and the user's accuracy was from 60% to 80% with SVM having the lowest accuracy. The least dominating classes in the Lindani valley bottom wetland are water and bare surface with areas ranging of 0.3- 5 ha for bare surface and 0.2 - 10 ha for water. The producer's accuracies for these two classes ranged between 0% and 75% for water and between 0% and 30% for bare surface. The SVM algorithm achieved 0% producer's accuracy for both the classes. The user's accuracy for the two classes was from 0% to 75% for bare surface and 0% to 50% for water. The SVM algorithm had a 0% user’s accuracy for both classes (Figure 6). The dominating landcover class in the Nylsvley floodplain is bare surface with an estimated area ranging from 362 ha to 495 ha. The CART algorithm estimated the lowest area. The least dominating class is long grass comprising mainly common reeds species such as *Phragmites australis*. The area for the least dominating wetland cover class ranged between 130 ha and 352 ha. The highest area was estimated by CART, and the lowest by NB. The producer’s accuracy for the dominating class ranged between 62% and 88% with the highest being attained by the NB and the lowest by the SVM model. The producer’s accuracy for the least dominating class ranged between 33% and 66.6% with the highest attained by NB model and lowest by CART and RF models. The user’s accuracy for the most dominating class ranged between 22% and 71% with highest recorded by SVM and CART and the lowest by the NB model. User’s accuracy for the least dominating ranged between 20% and 66.6% with highest recorded by RF and lowest by CART.

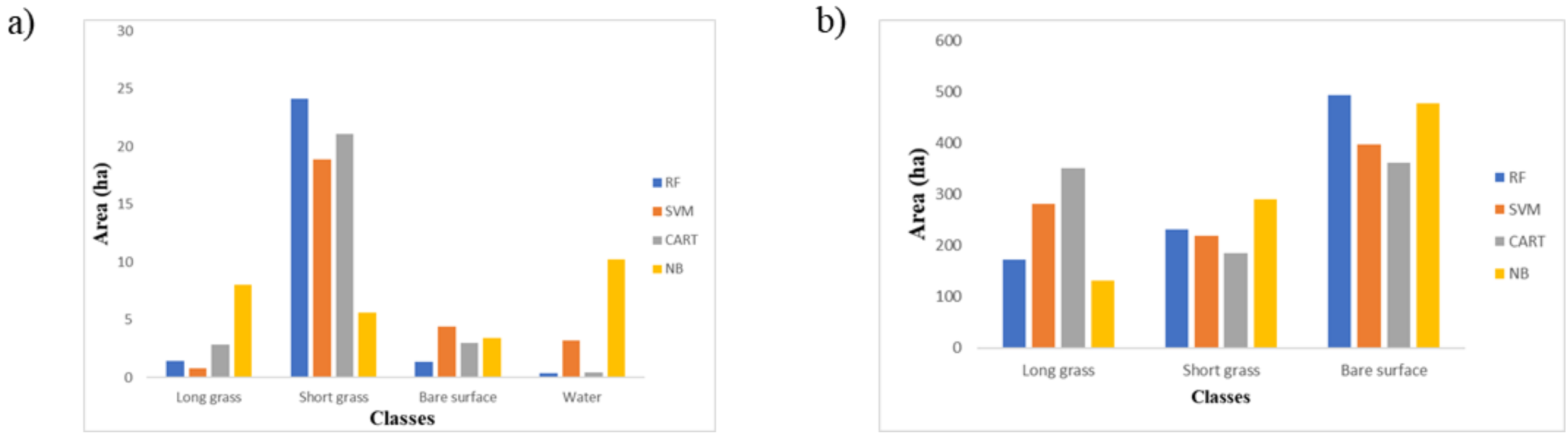

**Figure .5** Wetland cover class areas for a) Lindani valley bottom and b) Nylsvley floodplain

**Lindani valley bottom**

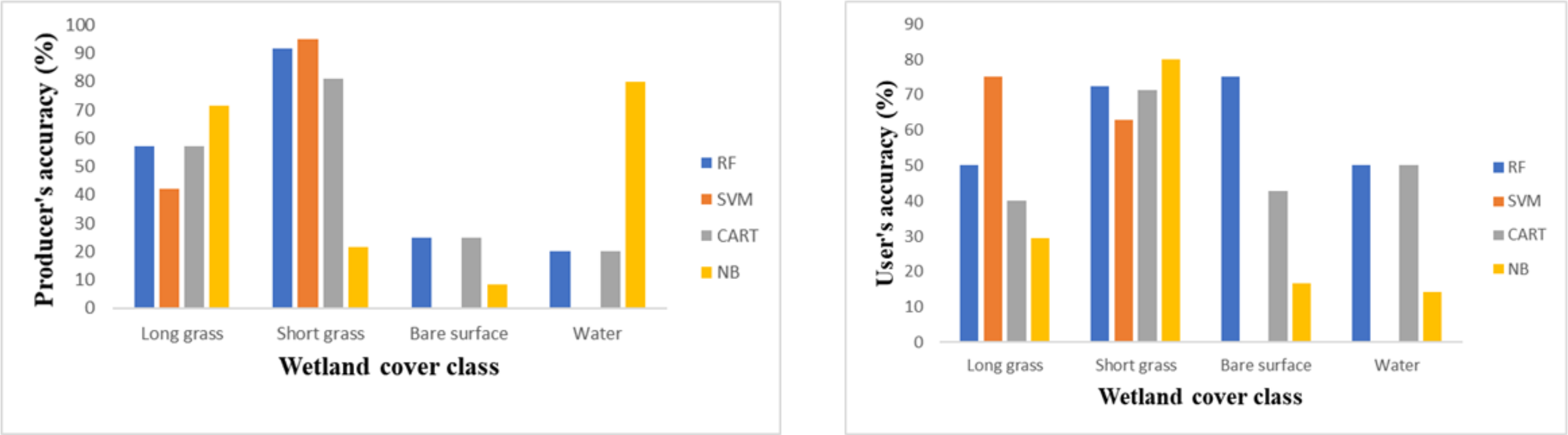

**Nylsvley floodplain**

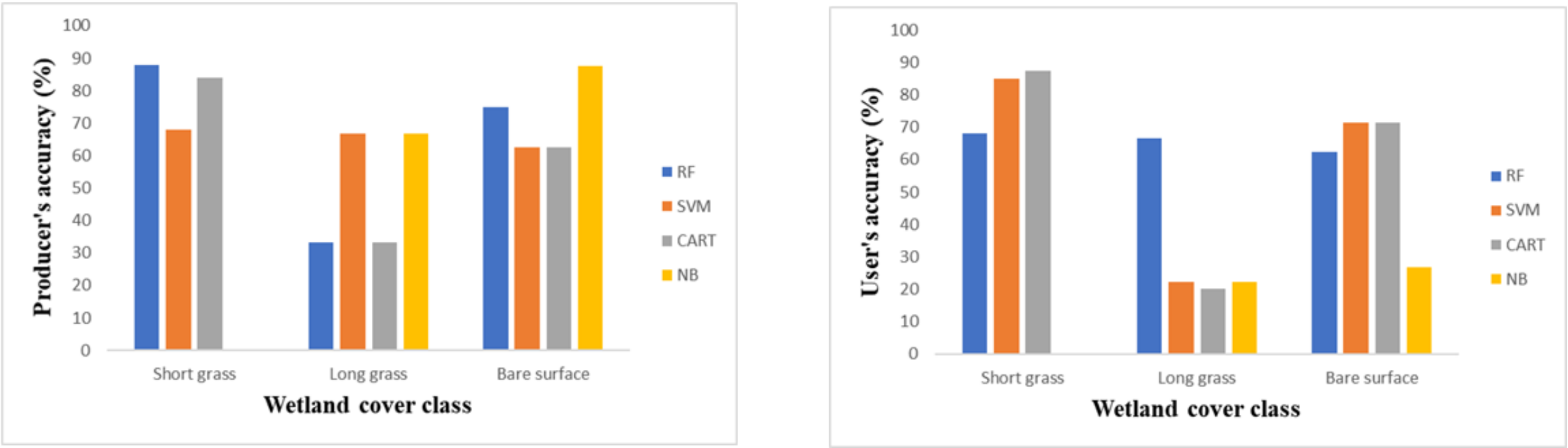

**Fig .6** Producer’s and user’s accuracy for the two wetlands classes

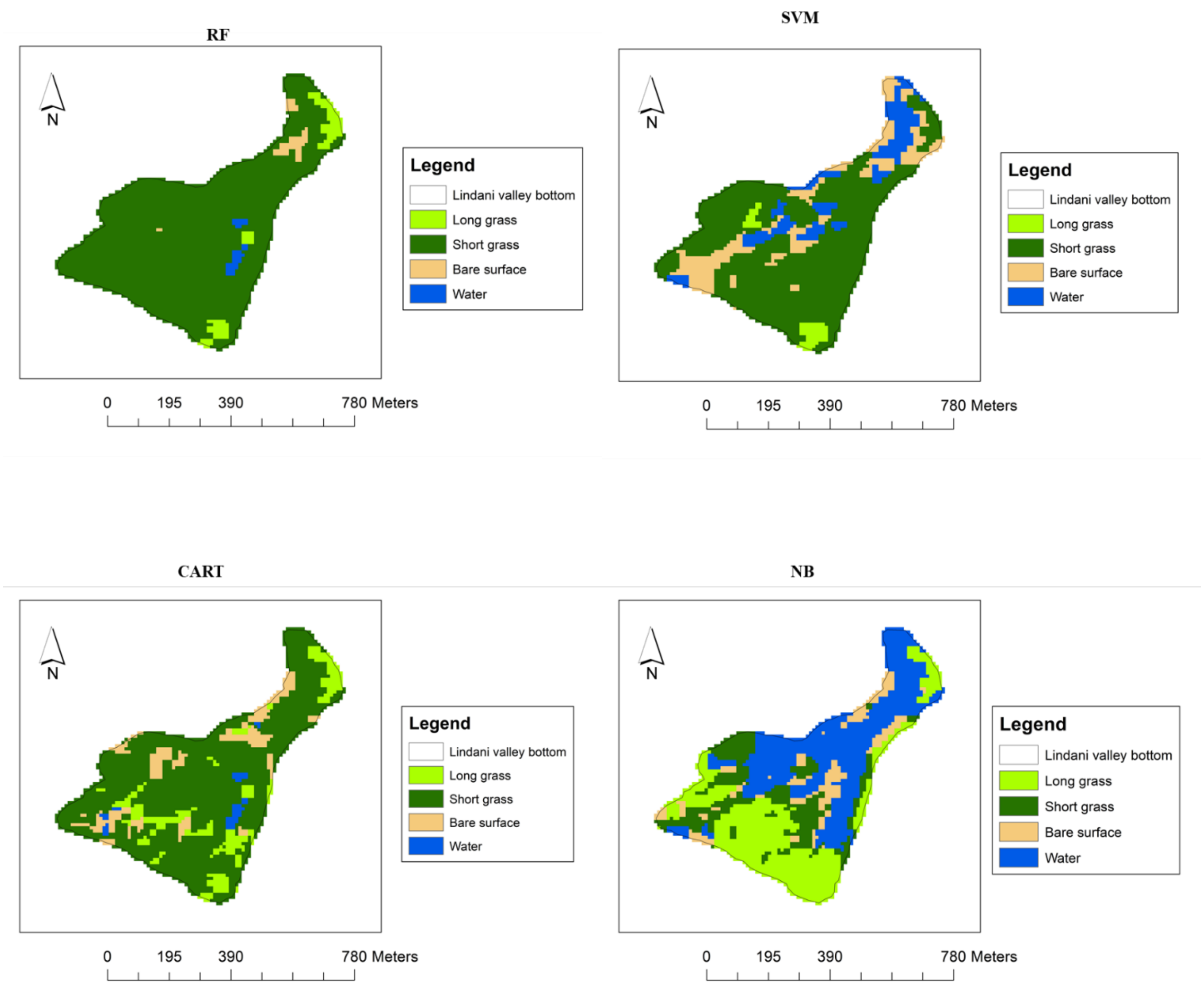

**Fig. 7:** Lindani valley bottom Sentinel 2 derived wetland cover classes based on the four used algorithms

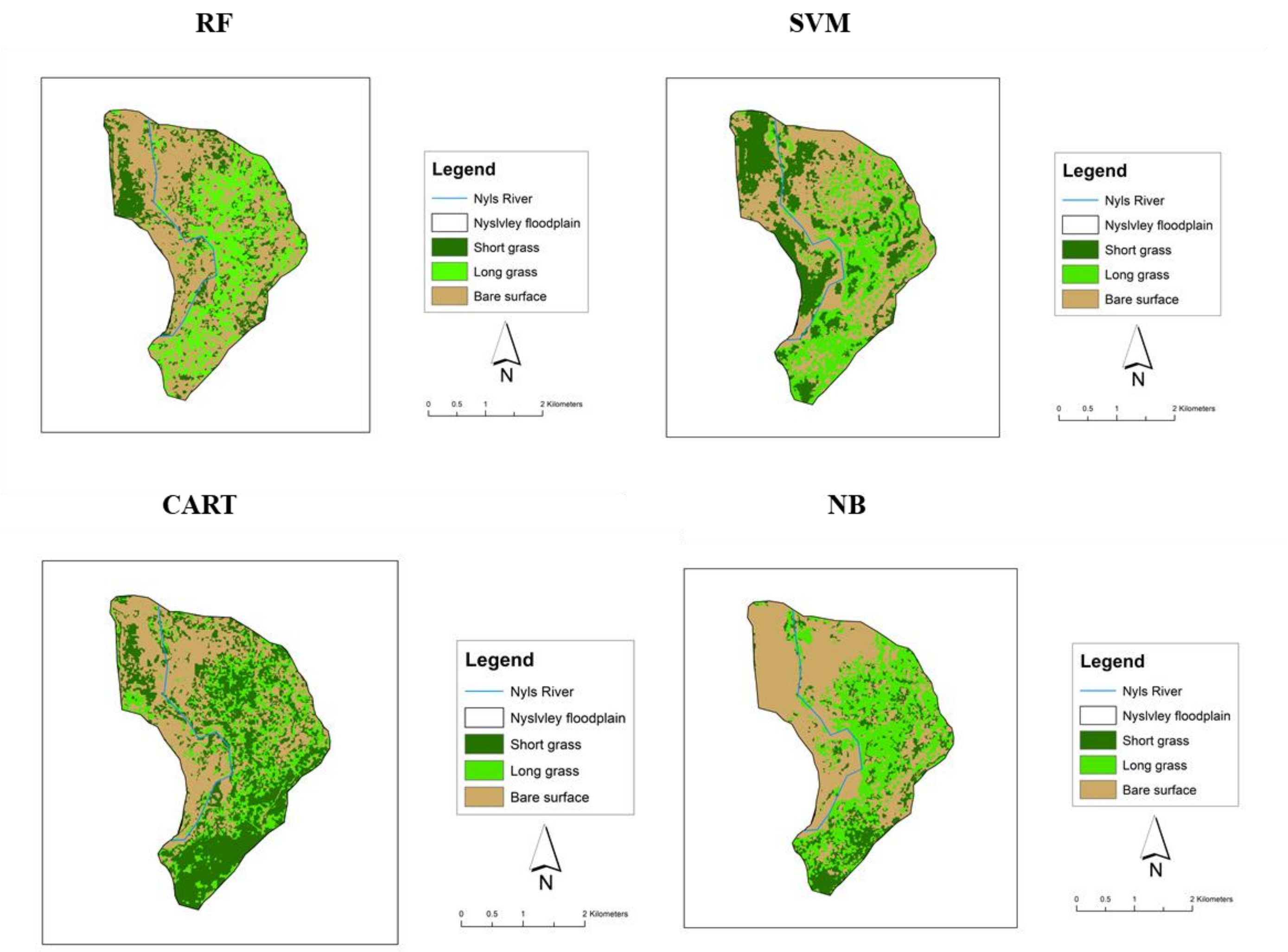

**Fig .8:** Nylsvley Floodplain Sentinel 2 derived wetland cover classes based on four algorithms used

## 5. Discussion

Accurate detection and monitoring of small seasonal and heterogeneous wetlands in the semi-arid regions is important for understanding the ecohydrological dynamics of these systems, as most of these wetlands particularly in the sub-Saharan Africa offer socio-economic benefits to the surrounding communities (Kabii and Kabii, 2005; Gardner *et al.*, 2009; Thamaga, Dube and Shoko, 2021). Advancements in data analytic tools provide unique opportunities to improve the detection and monitoring of semi-arid wetlands of variable sizes, which were not feasible, using the traditional remote sensing techniques. The introduction of cloud computing platforms such as Google Earth engine (GEE), offer great advantages such as advanced machine learning algorithms and parallel processing, memory efficiency and fast image processing power. The study sought to assess the capabilities of GEE cloud-computing platform in characterising and mapping semi-arid seasonal flooded wetlands at site specific scale, as well as suggesting a suitable GEE machine learning algorithm for characterizing and mapping such systems. In general, the results demonstrate the capabilities of Google Earth engine platform in characterising and mapping the semi-arid wetland systems of variable sizes with acceptable overall accuracies. In addition, RF, CART and SVM algorithms proved to be superior to the NB model with low OA values for both studied wetlands. Although higher OA were obtained using the latter algorithms, low producer's and user's accuracies occurred in the Lindani valley bottom for Bare and Water classes, especially using SVM model. The contributing factors to the low producer's and user's accuracies for the two classes were few training and validation sample points representing these classes due to less than a pixel spatial coverage of each class within the wetland boundary. Fewer training and validation points tend to reduce the level of accuracy during classification (Zhen *et al.*, 2013; Corcoran *et al.*, 2015; Mahdianpari *et al.*, 2020). In addition to few training and validation data points, the multi-year images used during the computation of median composite did not consider seasonality and yearly differences and that may have reduced the producer's and user's accuracies. The study by Noi Phan, *et al*., (2020) analysed the impact of different composition methods as well as input images on the classification accuracy of different landcover classes using GEE. The results showed that temporal aggregation considered during median compositing produced high accuracy values for the classification outputs. This therefore demonstrate the significance of temporal aggregations during the median compositing stage. Class spectral separability results show that the synergic use of all spectral feature significantly increase the separability of different classes for both wetlands with JM distances above 1.9. However, in the NIR region the logical expectation was that vegetation, water will have strong reflectance and adsorption in the NIR region, but in the case of Lindani valley bottom, water had stronger reflectance in this spectral region. This could be due to the submerged and floating wetland vegetation that interfere with the water signals thus causing stronger reflectance of water at NIR region (de Vries *et al.*, 2017). In addition, materials at the bottom of the shallow waters are known to affect absorption and reflectance of light by shallow waters (Vinciková, *et al.*, 2015). This could also be the case for this wetland with shallow water. Jones (2015), reported increased errors in mapping the spatial extent of water in areas within the greater Everglades where vegetation is floating. This showed the problem of discriminating between water and vegetation in such wetlands. In the study presented in this paper, both wetlands had short and long grass classes with higher reflectance values in the visible red-light region indicative of water stressed vegetation. Water stress vegetation is known to have stronger reflectance signals in visible red-light region (Artiola *et al.*, 1975). The study by Caturegli *et al.*, (2020) assessed the effects of water stress on spectral reflectance of Bermudagrass *(Cynodon dyctolon)* under controlled laboratory conditions. The results showed an increase in red light reflectance with increasing water stress, thus proving that water stress vegetation has a stronger reflectance in the visible red-light region. The indices were found to be the least useful in separating between the class for both wetlands, partly because seasonality and yearly differences were not considered during median image compositing of the selected images. Maximum NDVI values tends to correspond to the growing season in most cases. A study by Wang *et al.*, (2020) examined the response of the maximum NDVI values to  precipitation occurring during the period of active growth from 2000 to 2013 in the Alpine grassland site of the Tibetan Plateau. The results showed a positive linear relationship between precipitation and the maximum NDVI thus proving that NDVI is most useful during the peak growing season.

The findings of this study prove that the GEE platform and advanced machine learning algorithm have the potential to improve the detection and monitoring of small seasonal flooded wetlands in semi-arid regions using Sentinel-2 multi-year composite image. This has been previously a daunting task, using the conventional mapping methodologies and optical data. In addition, the results demonstrate that the most detectable wetland features were mostly wetland vegetation communities, although there were some challenges relating to accuracy particularly for the Lindani valley bottom system. The study provides baseline information and new insights about better enhancing small seasonal flooded wetlands from optical data at reasonable accuracy and moderately high-resolution, thus underscoring the significance of freely available optical data in monitoring semi-arid seasonal flooded systems. This is important for semi-arid regions with limited data access particularly sub-Saharan Africa where less attention is given to these systems due to limited information regarding their status although serving as important source of water for most communities. The findings also contribute towards the ongoing global wetland monitoring programmes such as Wetland Monitoring and Assessment Services for Transboundary Basins in Southern Africa (WeMAST) funded by European Union- Africa Global Monitoring for Environmental Security (EU Africa GMES). This programme aims at developing an integrated platform for wetlands assessment and monitoring that will support sustainable management in transboundary basins. Furthermore, the study further reports to the sustainable development goal 6.6, seeking to halt degradation and destruction of ecosystems including wetlands and assist in recovery of the already degraded systems.

## 6. Conclusion and Recommendations

The current study was aimed at characterising and mapping two seasonal flooded wetlands in the Limpopo Transboundary River Basin, with the objective of assessing the usefulness of GEE cloud computing platform in producing maps of the studied wetlands as well as suggesting possible GEE algorithms for detecting and mapping such systems. The main findings indicate the capabilities of GEE in mapping seasonal semi-arid wetlands system of variable size with reasonable overall accuracies, and RF CART and SVM algorithms being superior to the NB model. Although reasonable overall accuracies were attained, there were poor producer’s and user’s accuracies for some classes such as Water and Bare surface especially for the Lindani valley bottom wetland. This can be attributed to less than a pixel spatial coverage of these classes within wetlands perimeter, thus resulting in difficulties in their detection to the highest precision using Sentinel-2 composite data. In addition, the seasonality and yearly difference were not considered which could have significantly affected the results because some features such as water tend to correspond to seasonality changes, especially for semi-arid season. It is therefore recommended that temporal variability be considered in-order to capture the peak growing season of the systems and thus better enhancing the wetland features. Spectral confusions were also observed between water and some wetland vegetation, which resulted in the higher reflectance of water in the NIR region. In avoiding such, the study recommends the integration of Synthetic Aperture Radar (SAR) data to the optical data since SAR data can penetrate through forested vegetation thus minimising the effect of floating vegetation on the detection of water class.

## Declaration of competing interest

The authors declare that they have no known competing financial interests or personal relationships that could have appeared to influence the work reported in this paper.

## Acknowledgments

The authors would like to thank the anonymous reviewers for their valuable input to this paper as well as the Department of Earth Science, University of the Western Cape students Mr Eugene Sagwati Maswanganye and Ms Tatenda Dzurume for assisting with field data collection. The authors would also like to thank the Department of Economic Development, Environment and Tourism in Limpopo and the owners of the Lindani Private Game Lodge for granting access to the studied wetlands.

## Funding

The study was funded by the South African National Research Foundation (NRF) and The Global Monitoring funded this research for Environment and Security (GMES)—Africa through the WeMAST Project.